\documentclass[twocolumn]{aastex63}

\received{TBD}
\accepted{TBD}
\acceptjournal{ApJ}

\submitjournal{ApJ}

\shorttitle{Red galaxies at $z\sim12$}
\shortauthors{Mitsuhashi et al.}

\graphicspath{{./}}
\renewcommand\baselinestretch{0.92}
\usepackage{amsmath}
\tabletypesize{\footnotesize}
\usepackage{multirow}

\usepackage{ulem} 
\def\blue#1 {{\textcolor{blue}{#1}}\ }

\def\kms{\,km\,s$^{-1}$}

\def\red#1 {{\textcolor{red}{#1}}\ }

\begin{document}

\title{Discovery of red galaxy candidates at $z\sim12$: \\Early dust growth or significant nebular emission with high-temperature stars?}

\correspondingauthor{Ikki Mitsuhashi}
\email{ikki0913astr@gmail.com}

\author[0000-0001-7300-9450]{Ikki Mitsuhashi}
\affiliation{Department for Astrophysical \& Planetary Science, University of Colorado, Boulder, CO 80309, USA}

\author[0000-0002-1714-1905]{Katherine A. Suess}
\affiliation{Department for Astrophysical \& Planetary Science, University of Colorado, Boulder, CO 80309, USA}

\author[0000-0001-6755-1315]{Joel Leja}
\affiliation{Department of Astronomy \& Astrophysics, The Pennsylvania State University, University Park, PA 16802, USA}
\affiliation{Institute for Computational \& Data Sciences, The Pennsylvania State University, University Park, PA 16802, USA}
\affiliation{Institute for Gravitation and the Cosmos, The Pennsylvania State University, University Park, PA 16802, USA}

\author[0000-0001-8460-1564]{Pratika Dayal}
\affiliation{Kapteyn Astronomical Institute, University of Groningen, 9700 AV Groningen, The Netherlands}
\affiliation{Canadian Institute for Theoretical Astrophysics, 60 St George St, University of Toronto, Toronto, ON M5S 3H8, Canada}
\affiliation{David A. Dunlap Department of Astronomy and Astrophysics, University of Toronto, 50 St George St, Toronto ON M5S 3H4, Canada}
\affiliation{Department of Physics, 60 St George St, University of Toronto, Toronto, ON M5S 3H8, Canada}

\author[0000-0002-1109-1919]{Robert Feldmann}
\affiliation{Department of Astrophysics, University of Zurich, CH-8057, Switzerland}

\author[0000-0001-7201-5066]{Seiji Fujimoto}
\altaffiliation{NHFP Hubble Fellow}
\affiliation{David A. Dunlap Department of Astronomy and Astrophysics, University of Toronto, 50 St George St, Toronto ON M5S 3H4, Canada}
\affiliation{Department of Astronomy, The University of Texas at Austin, Austin, TX 78712, USA}

\author{Harley Katz}
\affiliation{Department of Astronomy \& Astrophysics, University of Chicago, 5640 S Ellis Avenue, Chicago, IL 60637, USA}

\author[0000-0003-2804-0648 ]{Themiya Nanayakkara}
\affiliation{Centre for Astrophysics and Supercomputing, Swinburne University of Technology, PO Box 218, Hawthorn, VIC 3122, Australia}

\author{Desika Narayanan}
\affiliation{Department of Astronomy, University of Florida, 211 Bryant Space Sciences Center, Gainesville, FL 32611 USA}
\affiliation{Cosmic Dawn Center (DAWN), Niels Bohr Institute, University of Copenhagen, Jagtvej 128, København N, DK-2200, Denmark}

\author[0000-0002-0108-4176]{Sedona H. Price}
\affiliation{Space Telescope Science Institute (STScI), 3700 San Martin Drive, Baltimore, MD 21218, USA}   
\affiliation{Department of Physics and Astronomy and PITT PACC, University of Pittsburgh, Pittsburgh, PA 15260, USA}  

\author[0000-0003-1614-196X]{John R. Weaver}
\affiliation{Department of Astronomy, University of Massachusetts, Amherst, MA 01003, USA}

\author[0000-0003-2919-7495]{Christina C.\ Williams}
\affiliation{NSF’s National Optical-Infrared Astronomy Research Laboratory, 950 North Cherry Avenue, Tucson, AZ 85719, USA}

\author[0000-0002-2057-5376]{Ivo Labbe}
\affiliation{Centre for Astrophysics and Supercomputing, Swinburne University of Technology, Melbourne, VIC 3122, Australia}

\author[0000-0001-5063-8254]{Rachel Bezanson}
\affiliation{Department of Physics and Astronomy and PITT PACC, University of Pittsburgh, Pittsburgh, PA 15260, USA}

\author[0000-0002-7570-0824]{Hakim Atek}
\affiliation{Institut d’Astrophysique de Paris, CNRS, Sorbonne Universit\'e, 98bis Boulevard Arago, 75014, Paris, France}

\author[0000-0003-2680-005X]{Gabriel Brammer}
\affiliation{Cosmic Dawn Center (DAWN), Niels Bohr Institute, University of Copenhagen, Jagtvej 128, K{\o}benhavn N, DK-2200, Denmark}

\author[0000-0002-7031-2865]{Sam E. Cutler}
\affiliation{Department of Astronomy, University of Massachusetts, Amherst, MA 01003, USA}

\author[0000-0001-6278-032X]{Lukas J. Furtak}\affiliation{Physics Department, Ben-Gurion University of the Negev, P.O. Box 653, Be’er-Sheva 84105, Israel}

\author[0000-0002-9651-5716]{Richard Pan}\affiliation{Department of Physics and Astronomy, Tufts University, 574 Boston Ave., Medford, MA 02155, USA}

\author[0000-0001-9269-5046]{Bingjie Wang
}
\altaffiliation{NHFP Hubble Fellow}
\affiliation{Department of Astrophysical Sciences, Princeton University, Princeton, NJ 08544, USA}

\author[0000-0001-7160-3632]{Katherine E. Whitaker}
\affiliation{Department of Astronomy, University of Massachusetts, Amherst, MA 01003, USA}
\affiliation{Cosmic Dawn Center (DAWN), Denmark}

\author{the UNCOVER/MegaScience team}

\begin{abstract}
We report the discovery of two $z\sim12$ galaxy candidates with unusually red UV slopes ($\beta_{\rm UV}\gtrsim-1.5$), and probe the origin of such colors at cosmic dawn. 
From \texttt{Prospector} fits to the UNCOVER/MegaScience dataset -- deep JWST/NIRCam imaging of Abell 2744 in 20 broad- and medium-bands -- we identify several new $z>10$ galaxies. 
Medium-band data improve redshift estimates, revealing two lensed ($\mu\sim3.3$) $z\sim12$ galaxies in a close pair with $\beta_{\rm UV}\gtrsim-1.5$ at an UV absolute magnitude of $M_{\rm UV}\sim-19\,{\rm mag}$, lying away from typical scatter on previously known $M_{\rm UV}$–$\beta_{\rm UV}$ relations.
SED fitting with \texttt{Prospector}, \texttt{Bagpipes}, and \texttt{EAZY} support their high-$z$ nature, with probability of low-$z$ interlopers of $p(z<7)<10\%$. 
The potential low-$z$ interlopers are $z\sim3$ quiescent galaxies (QGs), but unexpected to be detected at the given field of view unless $z\sim3$ QG stellar mass function has a strong turn up at $\log M_{\ast}[M_{\odot}]\sim9$.
Unlike typical blue high-redshift candidates ($\beta_{\rm UV}\lesssim-2.0$), these red slopes require either dust or nebular continuum reddening. The dust scenario implies $A_V\sim0.8\,{\rm mag}$, which is larger than theoretical predictions, but is consistent with a dust-to-stellar mass ratio ($\log M_{\rm dust}/M_{\ast}\sim-3$).
The nebular scenario demands dense gas ($\log n_{\rm H}\,[{\rm cm}^{-3}]\sim4.0$) around hot stars ($\log T_{\rm eff}\,[{\rm K}]\sim4.9$).
Spectroscopic follow-up is essential to determine their true nature and reveal missing galaxies at the cosmic dawn.
\end{abstract}




\keywords{galaxies: evolution - galaxies: formation - galaxies: high-redshift}

%
%
%
%
%
%
\section{Introduction}
Understanding galaxy formation and evolution from primordial galaxies to the present time is one of the main goals of modern astronomy.
Traditionally, early galaxies, especially at $z>4$, are selected by capturing their redshifted Lyman-$\alpha$ break at the rest-frame $1216\text{\AA}$ in the observed-frame optical \citep[e.g.,][]{1996ApJ...460L...1L,1996MNRAS.283.1388M,1999ApJ...519....1S,2007ApJ...670..928B,2015ApJ...803...34B,2010ApJ...725L.150O,2013ApJ...773...75O,2015ApJ...810...71F} and near-infrared (NIR) filters \citep[e.g.,][]{2022ApJ...929....1H,2022ApJ...940L..14N,2022ApJ...938L..15C,2023MNRAS.523.1036B,2023ApJS..265....5H,2023MNRAS.519.1201A}.
To avoid the contamination of low-$z$ interlopers, the selection criteria include a limitation to the rest-frame UV color and are biased to galaxies with blue UV colors \citep[e.g.,][]{2014ApJ...793..115B}.
Therefore, these typical and widely used selection criteria may miss intrinsically red high-$z$ galaxies.
Recent {\it James Webb Space Telescope} \citep{2023PASP..135f8001G} observations identified more than $30$ of $z>10$ galaxies, they mostly exhibit very blue rest-frame UV color \citep[$\beta_{\rm UV}<-2.0$, e.g., ][]{2024Natur.633..318C,2024ApJ...972..143C,2024arXiv240410751A,2025ApJ...983L..22K}, potentially indicating a lack of dust, dominant nebular continuum, or an active AGN at cosmic dawn. 

Cosmic dust originates from stellar phenomena, such as the ejecta of asymptotic giant branch (AGB) stars and rapidly cooling SN ejecta \citep{1980ApJ...239..193D,2001MNRAS.325..726T,2007ApJ...666..955N}, and grows in the dense interstellar media (ISM) through metal accretion and condensation into dust grains \citep{1990ASPC...12..193D,2009ASPC..414..453D,2011EP&S...63.1027I}.
As the age of the Universe is limited at high redshift, the dust content, especially within galaxies at the early cosmic age, is a direct indicator of the production/growth mechanisms \citep{2017MNRAS.471.3152P,2019MNRAS.490..540L,2022MNRAS.512..989D,2025ApJ...985L..21M,2025arXiv250918266N}.
Efforts to date have failed to detect dust continuum for galaxies at $z>8.5$ \citep{2023MNRAS.519.5076B,2023ApJ...950...61Y,2025arXiv250119384M}, although these dust non-detections are not entirely unexpected based on their blue colors and typical IRX-$\beta_{\rm UV}$ relation at $z\lesssim8$ \citep[e.g.,][]{2020A&A...643A...4F,2024ApJ...971..161M,2024MNRAS.527.5808B}.
On the other hand, some analytical models predict that these high-redshift blue galaxies should go through an obscured starburst phase \citep[e.g.,][]{2023MNRAS.522.3986F}.

Reddening in $z>9.5$ galaxies also arise from significant nebular continuum emission \citep{2024arXiv241114532S}, which can produce UV slopes as red as $\beta_{\rm UV}\sim-1.3$ \citep{2024arXiv240803189K,2025ApJ...982....7N}.
Reproducing $\beta_{\rm UV}>-1.5$ requires stellar effective temperatures of $T_{\rm eff}\gtrsim50000\,{\rm K}$ and gas density of $n_{\rm H}\gtrsim10^4\,{\rm cm}^{-2}$, which is significantly higher than the typical electron density \citep[e.g.,][]{2021ApJ...909...78D,2023ApJ...956..139I} and stellar temperature ($\sim35000\,{\rm K}$ in O-type stars, \citealt[e.g.,][]{2005A&A...436.1049M}, while sometimes O-type stars show up to $T_{\rm eff}\sim50000\,{\rm K}$, e.g.,  \citealt{2006ApJ...638..409H}).
These conditions are consistent with the existence of the very massive stars (VMS) or metal-poor stars, which are expected to have very high $T_{\rm eff}$ \citep{2002A&A...382...28S,2025A&A...693A.271S}.
The metal-poor star formation site, as represented by pop-{\sc iii} stars/galaxies, has long been of great interest \citep[e.g.,][]{1984ApJ...277..445C,2002ApJ...567..532H,2011MNRAS.415.2920I,2011ApJ...740...13Z}, and some metal-poor galaxy candidates have been recently identified \citep{2025arXiv250111678F,2025arXiv250710521M}.
However, these candidates are still limited to $z\sim6$ due to the challenges in covering the rest-frame optical wavelength and H$\alpha$ emission line at higher redshift.

In summary, the red galaxies at $z>10$ represent an exciting population of galaxies from two competing physical explanations: significant dust-reddening due to early dust growth in a metal-enriched galaxy, or nebular continuum-dominated, young, metal-poor star formation.
Specifically, the existence of metal-poor star-forming galaxies is expected at such an early epoch from theoretical studies \citep[e.g.,][]{2023MNRAS.522.3809V}.
However, to date, there are few robust red galaxy candidates at $z>10$.
In this paper, we present two plausible $z\sim12$ candidates with red rest-frame UV colors, as well as comparison $z>10$ galaxy candidates selected from the 20 NIRCam observations taken in the UNCOVER/MegaScience program.
We discuss three possibilities to explain the 20-band photometry of these objects.

The paper is organized as follows: Section \ref{sec:data} provides an overview of the datasets and analysis to infer the redshift of the two red galaxy candidates at $z\sim12$. Section \ref{sec:discussion} describes our results and discusses possible explanations for these red colors at $z\sim12$. The summary and future prospects are presented in Section \ref{sec:summary}. Throughout this paper, we assume a flat universe with the cosmological parameters of $\Omega_{\rm M}=0.3$, $\Omega_{\Lambda}=0.7$, $\sigma_{8}=0.8$, $H_0=70$ \kms ${\rm Mpc}^{-1}$, and \citet{2001MNRAS.322..231K} initial mass function (IMF).

%
%
%
%
%
%
\section{Two $\lowercase{z}\sim12$ red galaxy candidates}\label{sec:data}

\subsection{Data}\label{subsec:uncover}
We utilized the JWST observations of the Abell 2744 cluster field from the UNCOVER \citep{2024ApJ...974...92B,2024ApJS..270....7W} and MegaScience \citep{2024ApJ...976..101S} surveys.
Specifically, we used the UNCOVER/MegaScience DR3 photometric catalog containing 20 NIRCam broad- and medium-band filters.
The DR3 catalog includes additional imaging taken by other JWST programs in the Abell 2744 field, such as GLASS-ERS \citep[PI: Treu][]{2022ApJ...935..110T}, ALT \citep[PI: Naidu\&Matthee][]{2024arXiv241001874N}, and MAGNIF \citep[PI: Sun][]{2023arXiv231009327L}, as well as several GO programs (GO-3538, GO-2754; \citealt{2024MNRAS.528.7052C}).
Please refer to \citet{2024ApJ...976..101S} for more details on the DR3 catalog.

We selected high-redshift candidates in Abell2744 using the \texttt{Prospector}-$\beta$ model \citep{2023ApJ...944L..58W,2024ApJS..270...12W}, which recovers redshifts well \citep[$\sigma_{\rm NMAD}\sim0.15$,][]{2024ApJ...976..101S}.
We limited the parent sample to the sources covered by all medium-bands to avoid the contamination of low-$z$ emission line galaxies \citep[e.g.,][]{2023ApJ...943L...9Z,2023Natur.622..707A}.
To select the most secure candidates, we adopted the following additional criterion: (1) $z_{\rm best}>10$, where $z_{\rm best}$ is the peak value in $p(z)$ (2) robust photometric measurements ({\sc use\_phot}=1 in DR3 catalog), (3) ${\rm S/N}>5$ in LW bands (F277W, F356W, and F444W) and (4) $p(z<7)<5\%$.
After excluding clear artifacts in visual classification, we found seven galaxies satisfying the criteria above (Table \ref{tab:tab1}).
Our candidates includes two of four $z_{\rm spec}>10$ from NIRSpec/PRISM spectroscopy in the UNCOVER field \citep{2024ApJ...977..250F,2025ApJ...982...51P}.
The remaining two sources with $z_{\rm spec}>10$ \citep{2023ApJ...957L..34W} are not included in the sample.
UNCOVER-z12 is excluded because it lies outside the medium-band footprint, and UNCOVER-z13 is excluded because ${\rm S/N}_{\rm F277W} = 2.6$.

\begin{figure}[tb]
\begin{center}
\includegraphics[width=8.5cm,bb=0 0 1000 650, trim=0 1 0 0cm]{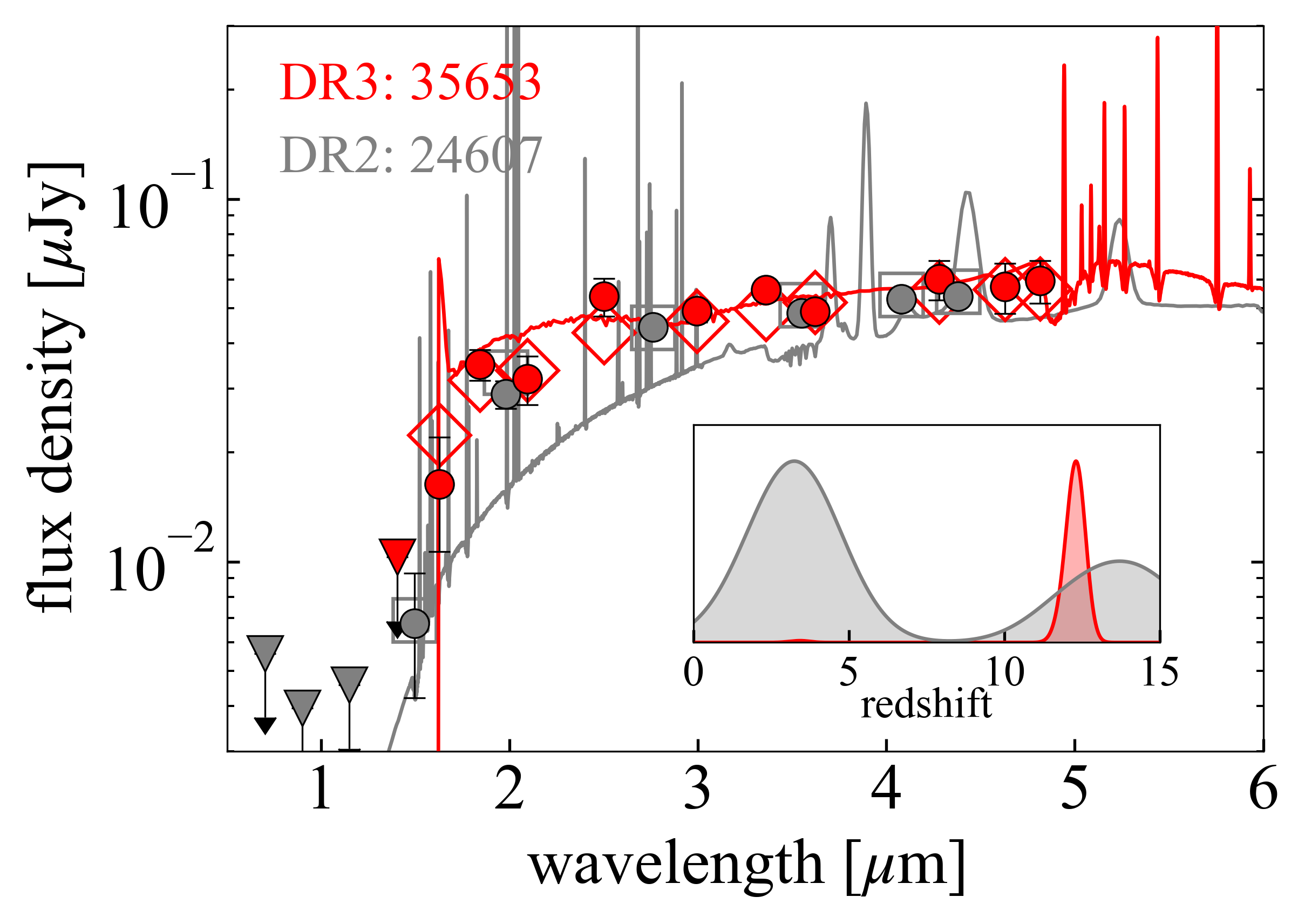}
\caption{Photometry of one of the target galaxies in UNCOVER DR2 (gray) and DR3 (red, with MegaScience data) with the best-fit \texttt{Bagpipes} SEDs. The inserted panel shows $p(z)$ based on the photometry in each catalog. The best-fit SED and $p(z)$ changes significantly before and after adding the photometric points of the medium bands.}
\label{fig:SED_DR3DR2}
\end{center}
\end{figure}

Among seven candidate galaxies at $z>10$, two are newly selected as the $p(z)$ significantly changed before and after adding medium-band observations.
Figure \ref{fig:SED_DR3DR2} demonstrates how the addition of medium-band photometry is critical to change $p(z)$ in one of the new candidates.
In the broad band photometry, these objects had a high probability solution of a strong line emitter at $z\sim5$. 
However, after densely sampling the SED via the addition of MegaScience medium bands, it is clear that this source does not host strong emission lines, and the low-$z$ probability nearly disappears.

%
%
%
%
%
%
%
%
%
%
%
%

%
%
%
%
%
%
\begin{deluxetable*}{ccccccccccc}
\tablecaption{Summary of $z>10$ candidates in UNCOVER/MegaScience \label{tab:tab1}}
\tablewidth{0pt}
\tablehead{\colhead{ID(DR3)$^{\small \dagger}$} & 
\colhead{RA} &
\colhead{Dec} &
\colhead{$z_{\rm phot}^{\small \ddag}$} &
\colhead{$z_{\rm spec}^{\small \ast}$} &
\colhead{$\mu$} &
\colhead{$M_{\rm UV}^{\small \P}$} & 
\colhead{$\beta_{\rm UV}^{\small \P}$} &
\colhead{$\log M_{\ast}^{\small \ddag}$} &
\colhead{ID(DR2)$^{\small \S}$} & 
\colhead{ID(MSA)$^{\small \S}$} \\
 &  &  &  &  &  & [mag] &  & [$M_{\odot}$] &  &
}
\startdata
20930 & 3.57098 & -30.40707 & $11.4_{-0.4}^{+0.4}$ & - & $2.260_{-0.002}^{+0.002}$ & $-18.69_{-0.10}^{+0.12}$ & $-1.78_{-0.21}^{+0.21}$ & $8.23_{-0.24}^{+0.30}$ & 11259 & - \\
25494 & 3.59500 & -30.40074 & $10.2_{-0.1}^{+0.5}$ & - & $13.213_{-0.030}^{+0.118}$ & $-17.67_{-0.03}^{+0.03}$ & $-2.57_{-0.06}^{+0.05}$ & $7.62_{-0.16}^{+0.06}$ & 15197 & - \\
35653 & 3.54085 & -30.38064 & $12.4_{-0.2}^{+0.1}$ & - & $3.270_{-0.002}^{+0.002}$ & $-18.96_{-0.04}^{+0.06}$ & $-1.33_{-0.08}^{+0.09}$ & $8.48_{-0.15}^{+0.18}$ & 24607 & - \\
35654 & 3.54080 & -30.38047 & $12.3_{-0.2}^{+0.2}$ & - & $3.317_{-0.003}^{+0.002}$ & $-19.13_{-0.04}^{+0.05}$ & $-1.54_{-0.08}^{+0.07}$ & $8.62_{-0.32}^{+0.34}$ & 24608 & - \\
38201 & 3.56707 & -30.37786 & $10.8_{-0.1}^{+0.1}$ & 10.071 & $4.109_{-0.003}^{+0.003}$ & $-19.02_{-0.04}^{+0.03}$ & $-2.43_{-0.08}^{+0.06}$ & $7.91_{-0.20}^{+0.16}$ & 27025 & 26185 \\
42089 & 3.51193 & -30.37186 & $10.3_{-0.5}^{+0.4}$ & - & $1.715_{-0.002}^{+0.002}$ & $-20.28_{-0.03}^{+0.04}$ & $-1.79_{-0.05}^{+0.05}$ & $9.09_{-0.40}^{+0.28}$ & 30818 & - \\
49514 & 3.59011 & -30.35974 & $10.9_{-0.2}^{+0.1}$ & 10.255 & $2.207_{-0.001}^{+0.001}$ & $-19.96_{-0.05}^{+0.05}$ & $-2.62_{-0.09}^{+0.09}$ & $8.26_{-0.15}^{+0.19}$ & 38095 & 37126  
\enddata

\tablecomments{
\vspace{-6pt}\tablenotetext{\small \dagger}{ID in the DR3 catalog \citep{2024ApJ...976..101S}.}
\vspace{-6pt}\tablenotetext{\small \ddag}{Photometric redshift and stellar mass derived by \texttt{Prospector} \citep{2023ApJ...944L..58W}.}
\vspace{-6pt}\tablenotetext{\small \ast}{Spectroscopic redshift measured in \citep[][see also, \citealt{2025ApJ...982...51P}]{2024ApJ...977..250F}.}
\vspace{-6pt}\tablenotetext{\small \P}{UV absolute magnitude and UV slope calculated in this work (Sec \ref{subsec:uncover}).}
\vspace{-6pt}\tablenotetext{\small \S}{IDs in the DR2 catalog and msa observations \citep{2024ApJ...974...92B,2024ApJS..270....7W}.}
\vspace{-22pt}
}
\end{deluxetable*}
%
%
%
%
%

\subsection{UV slope $\beta_{\rm UV}$ measurements}\label{subsec:slopemeasure}

We measured the UV slope $\beta_{\rm UV}$ by power-law fitting directly to the observed photometry \citep{2014MNRAS.440.3714R} at the rest-frame wavelength range of $1300$--$3200$\AA\ as in recent works \citep[e.g.,][]{2023MNRAS.520...14C,2024MNRAS.531..997C,2024arXiv240410751A,2024ApJ...965...98C}.
The $M_{\rm UV}$ and its uncertainties have been measured by averaging the power-law spectra at the rest-frame $1450$--$1550$\,\AA.
We also tested another approach, including emission line modeling using \texttt{Bagpipes}.
Here we assumed the delayed-$\tau$+burst SFH with $\tau$ ranging from 0--5\,Gyr plus a recent burst with variable timescale 1--100\,Myr \citep[see also][]{2024ApJ...965...98C}, the Calzetti dust attenuation law in the range of $A_V[{\rm mag}]=[0,5]$, stellar/gas metallicity in the range of $Z[Z_{\odot}]=[0.005,0.5]$ \citep{2023ApJS..269...33N,2024ApJ...973....8H,2024ApJ...972..143C,2025NatAs...9..155Z,2025A&A...695A.250A}, the ionization parameter $U_{\rm ion}=[-4,-2]$ and ionizing photon escape fraction $f_{\rm esc}=[0,1]$ to explore small $\beta_{\rm UV}$ regime \citep[i.e., $\beta_{\rm UV}<-2.6$;][]{2022ApJ...941..153T,2024arXiv241119893Y}.
We then perform a power-law fit of the best-fit \texttt{Bagpipes} model and confirm that the $\beta_{\rm UV}$ and $M_{\rm UV}$ derived from \texttt{Bagpipes} modeling yield consistent values with those obtained from simple power-law fittings.

\begin{figure}[htbp]
\begin{center}
\includegraphics[width=8.5cm,bb=0 0 1000 650, trim=0 1 0 0cm]{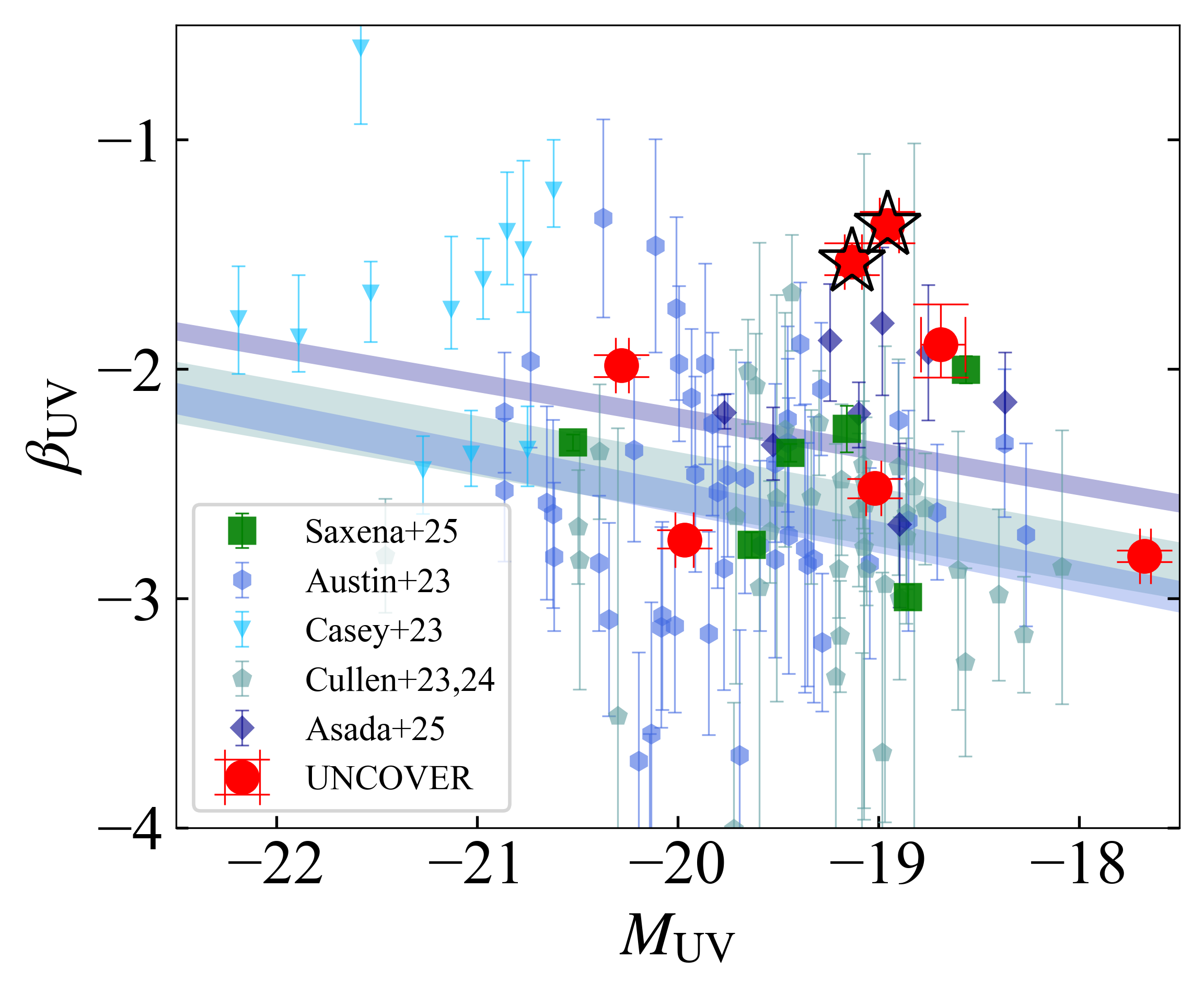}
\caption{UV absolute magnitude ($M_{\rm UV}$) vs UV slope $\beta_{\rm UV}$ of our $z>10$ galaxies (red) with previous photometric \citep[blue-ish markers,][]{2023MNRAS.520...14C,2024MNRAS.531..997C,2024arXiv240410751A,2024ApJ...965...98C,2025arXiv250703124A} and spectroscopic \citep[green markers][]{2024arXiv241114532S} constraints. The best-fit relationships from \citet{2024MNRAS.531..997C}, \citet{2024arXiv240410751A}, and \citet{2025arXiv250703124A} are shown in the corresponding color-shaded region. Our $z>10$ galaxies broadly align with the scaling relation from \citet{2025arXiv250703124A}, except for two galaxies surrounded by black stars.
}
\label{fig:MUV_beta}
\end{center}
\end{figure}

Figure \ref{fig:MUV_beta} shows our $\beta_{\rm UV}$ measurement as well as the recent photometric \citep{2023MNRAS.520...14C,2024MNRAS.531..997C,2024arXiv240410751A,2024ApJ...965...98C,2025arXiv250703124A} and spectroscopic \citep{2024arXiv241114532S} measurements at $z>10$.
Thanks to the dense sampling of the SED from all broad- and medium-band NIRCam filters, our constraints on $\beta_{\rm UV}$ are relatively tight with an average uncertainty of $\lesssim0.1$.
We found that five of seven galaxies at $z>10$ have UV slopes of $\beta_{\rm UV}\lesssim-2.0$, and align better with the relationship in \citet{2025arXiv250703124A} than others, suggesting redder colors on average than the other scaling relations \citep{2023MNRAS.520...14C,2024MNRAS.531..997C,2024arXiv240410751A}.
\citet{2025arXiv250703124A} utilized the photometric redshift inferred from NIRCam medium- and broad-band taken in CANUCS/Technicolor survey, whereas others mainly use broad-band photometry.
The alignment support that traditional Lyman break selection for fewer bands may miss some fraction of $z>10$ galaxies with intrinsically red UV slopes.

We find two galaxies (ID35653 and ID35654) which significantly deviate from every relationship with $\beta_{\rm UV}\sim-1.5$ at $M_{\rm UV}\sim-19\,{\rm mag}$, one of the reddest galaxies with the robust $\beta_{\rm UV}$ measurements.
Hereafter, we focus on the two red galaxy candidates at $z\sim12$ and investigate the origin of their red UV colors.

%
%
%
%
%
%

\subsection{Identification and photo-$z$ estimation}\label{subsec:main}

%
%
%
%
%
%
\begin{figure*}[htbp]
\begin{center}
\includegraphics[width=18cm,bb=0 0 1000 650, trim=0 1 0 0cm]{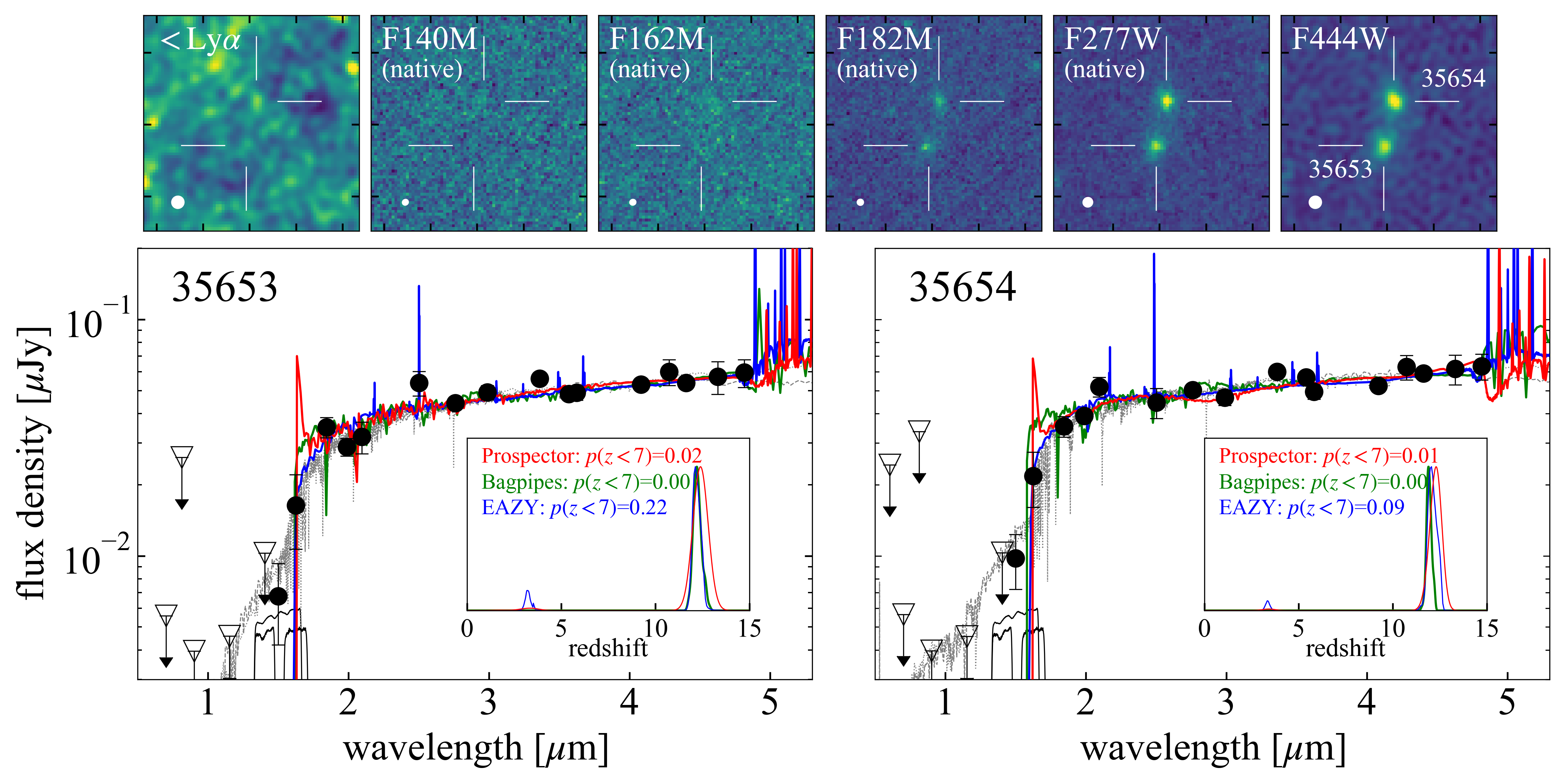}
\caption{All NIRCam photometry of two red $\beta_{\rm UV}$ galaxy candidates at $z\sim12$. 
The best-fit SEDs in \texttt{EAzY}, \texttt{Bagpipes}, and \texttt{Prospector}, and the SEDs of low-$z$ contaminant ($z\sim3$ QGs) obtained by forcing $p(z)$ to $z<7$ are shown in the colors and gray lines, respectively. Corresponding $p(z)$ and $p(z<7)$ are shown in the inserted panels in the bottom right. 
The top panels show thumbnails of the two galaxies in different wavelengths. From left to right, we show the stacked image of the bands shorter than Ly$\alpha$ at $z\sim12$, F140M, F162M, F277W image with original spatial resolution, and F444W.
}
\label{fig:SED_chi2}
\end{center}
\end{figure*}

Figure \ref{fig:SED_chi2} shows the SEDs of the two red galaxy candidates at $z\sim12$. 
As shown in the top panels, they are a pair of galaxies with a spatial separation of $0\farcs63/\mu$ ($\sim1.3\,{\rm kpc}$ at $z\sim12$ after lens magnification correction).
They are undetected in all bands shorter than F140M at the $<1.5\sigma$ level and exhibit a rapid decrease in flux (a factor of seven) between F182M and F115W, supporting their $z\sim12$ solution.
In addition to the existing \texttt{Prospector} fits, we performed additional SED fitting with \texttt{EAzY} \citep{2008ApJ...686.1503B} and \texttt{Bagpipes} \citep{2018MNRAS.480.4379C} to further quantify the probability of a low-redshift interloper solution.
We perform \texttt{EAzY} fits using the \texttt{agn\_blue\_sfhz\_13} template, which includes a redshift-dependent star formation history and dust attenuation as well as templates for several extreme objects discovered by JWST \citep{2024A&A...691A..52K}.
We perform \texttt{Bagpipes} fits assuming a delayed-$\tau$+burst SFH as mentioned above.
We allow the absolute magnitudes of V-band attenuation ($A_V$) with the Calzetti extinction curve to span 0--3\,mag and the metallicity to span 0.01--2\,$Z_{\odot}$ to capture a reasonable range.
The redshift ranges $z_{\rm phot}=0$--15, and the age of the galaxy has been capped at the age of the Universe at the given redshift. 
We fixed $\log U=-3.0$ and verified that this choice does not affect the photometric redshift measurements.

Figure \ref{fig:SED_chi2} shows the results of our SED fits with $p(z)$ in the inserted panels. 
Most of the redshift probability is dominated by the $z\sim12$ solution, with $p(z<7)<0-20\%$ depending on the SED fitting codes.
Thanks to the medium-band observations, the possibilities of the dusty line emitter at $z\sim4$--5 have been almost completely denied \citep[e.g.,][]{2023ApJ...943L...9Z,2023Natur.622..707A}.
There is no counterpart among the known line emitters identified by ALT \citep{2024arXiv241001874N} or dust continuum sources from ALMA observations \citep{2025ApJS..278...45F}, further supporting the fact that they are not line emitters.

The red colors of these sources do permit a small other $p(z)$ peak than $z\sim12$: $z\sim3$ quiescent galaxies with a strong Balmer break instead of a Lyman break. 
While permitted, this solution is not preferred by any of our three SED fitting codes because even a very strong Balmer break struggles to reproduce the $\sim7\times$ flux decrease we observe between F182M and F115W.
This is much larger than the most significant Balmer break allowed by SPS models from \texttt{fsps} between rest-frame U- and V-band (corresponding to F115W and F182M at $z\sim3$), which is $\sim4$ without dust extinction \citep{2024ApJ...974..145S}.
The best-fit SEDs forced to $z<7$ using \texttt{Bagpipes} indicate old ($t_{\rm age}\sim1\,{\rm Gyr}$) stellar populations with small dust extinction ($A_V\sim0.15$), which enlarges the strength of the break to $\sim5$.
Strong Balmer breaks are seen in the so-called ``Little Red Dots'' - for example, \citet{2025arXiv250316596N} reported a very strong break strength of $\sim8$ at $z\sim8$.
However, if the low-$z$ contaminants are the LRD at $z\sim3$, we expect strong H$\alpha$ emission at $\sim2.6\,\mu{\rm m}$ and point-source-like morphology, which the red galaxies do not exhibit.

We also calculated $\Delta\chi^{2}$ from \texttt{Bagpipes} and \texttt{EAZY} fitting as presented in several studies \citep[e.g.,][]{2023ApJS..265....5H,2024ApJ...965...98C}, where $\Delta\chi^{2}$ is difference of $\chi^{2}$ value between $z\sim12$ and $z\sim3$ solutions.
\texttt{Bagpipes} gives $\Delta\chi^{2}=8$--10, which is comparable with the usually adopted criterion $\Delta\chi^{2}>9$, while \texttt{EAZY} returns $\Delta\chi^{2}=2$--3.
Given the difficulties in obtaining clear answers from $\chi^{2}$ values of $z\sim12$ and $z\sim3$ solutions, future spectroscopic observations are necessary to confirm their redshifts.

\subsection{Number densities and sizes}\label{subsec:others}

%
%
%
%
%
%
\begin{figure*}[htbp]
\begin{center}
\includegraphics[width=18cm,bb=0 0 1000 650, trim=0 1 0 0cm]{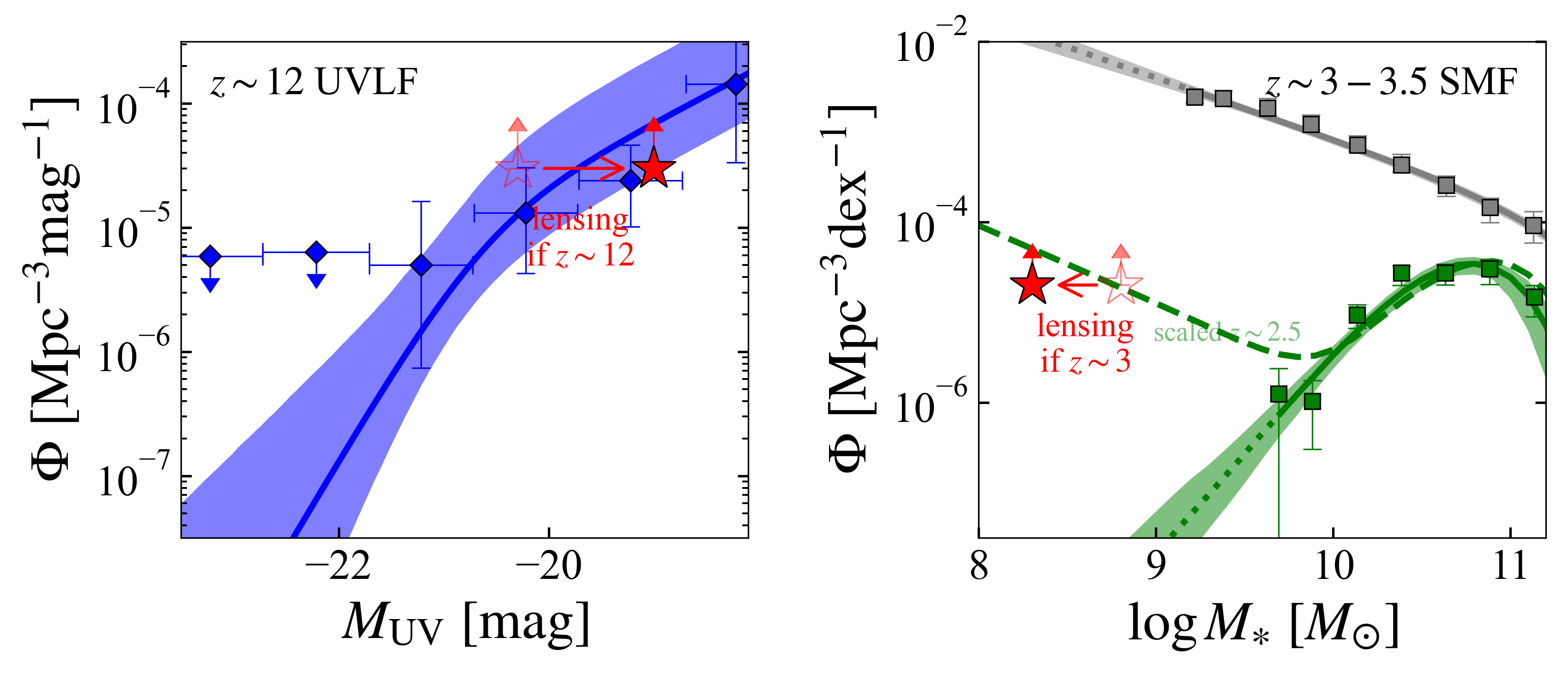}
\caption{(left) The number density of the $z\sim12$ LBGs \citep{2025ApJ...980..138H} and (right) $z\sim3$ QGs \citep[low-$z$ contaminant of the $z\sim12$ LBGs][]{2023A&A...677A.184W}. The magnitude of two red $\beta_{\rm UV}$ candidates before and after lensing correction is shown in open and filled stars, respectively. The intrinsically fainter galaxies are more advantageous than brighter ones, from the perspective of avoiding low-$z$ contaminants.}
\label{fig:Ncomp}
\end{center}
\end{figure*}

We examine the sizes and implied number densities of these objects to further explore whether a statistical view prefers a high- or low-redshift solution.
The lens-corrected K-band magnitude of the target galaxies is $m_{\rm K}\sim29\,{\rm mag}$ and the corresponding absolute UV magnitude assuming $z\sim12$ is $M_{\rm UV}\sim-19\,{\rm mag}$.
In the case of $z\sim3$ QGs, the lens-corrected stellar mass derived using \texttt{Bagpipes} and \texttt{eazy} is $\log M_{\ast}[M_{\odot}]\sim8.3$.
Figure \ref{fig:Ncomp} shows the UV LF at $z\sim12$ and the QG SMF at $z\sim3$ (see also,  \citealt{2023ApJ...955..130F,2025ApJ...980..138H}).
Given the UNCOVER/MegaScience field of view, the probability of detecting at least two $z\sim12$ galaxies at $M_{\rm UV}\sim-19$ based on \citet{2025ApJ...980..138H} UV LF is $>60\%$.
At the stellar mass of $\log M_{\ast}[M_{\odot}]\sim8.3$, there is no direct constraint on the QGs mass function at $z\sim3$ because of the lack of a survey area or depth.
Extrapolating QG SMF in \citet[][constrained down to $M_{\ast}\sim10^{9.6}M_{\odot}$]{2023A&A...677A.184W} results in detection probability of two $z\sim3$ QGs is $<10^{-5}\%$, while the probability goes up to $>80\%$ if we assume the same shape of QG SMF with $z\sim2.5$ \citep[][constrained down to $M_{\ast}\sim10^{8.7}M_{\odot}$]{2022ApJ...940..135S}.
If the two galaxies are actually $z\sim3$ QGs, that suggests a surprisingly large abundance of the low-mass QG at $z>3$, supporting a strong turnover at $\log M_{\ast}[M_{\odot}]\sim9.5$ due to environmental quenching \citep[e.g.,][]{2022ApJ...940..135S,2024arXiv241209592H}.
However, environmental quenching is not expected to be significant at $z>3$ \citep[e.g.,][]{2024ApJ...966L...2X}.
In summary, the $z\sim12$ solution is preferred from the perspective of the simple number density comparison.

As shown in Figure \ref{fig:SED_chi2}, these two galaxies are spatially resolved in JWST images with moderate gravitational magnifications ($\mu\sim3.3$).
S{\'e}rsic profile fitting using {\sc pysersic} \citep[][Zhang et al. in prep]{2024arXiv241206957M} indicate the effective radii of $0\farcs09_{-0.01}^{+0.01}$ and $0\farcs08_{-0.01}^{+0.01}$ in ID35653 and 35654, respectively, after lens magnification correction.
Here we applied uniform magnification correction in $\sqrt{3.3}=1.8$, since there is no significant difference between radial and tangential magnification factor \citep[$<10\%$,][]{2023MNRAS.523.4568F}.
These correspond to $0.32_{-0.04}^{+0.05}\,{\rm kpc}$ and $0.31_{-0.03}^{+0.04}\,{\rm kpc}$ in the physical scale at $z\sim12$, respectively. 
This fact confirms that they are not local dwarf stars.
Given that even $z\sim14$ galaxies are sometimes spatially extended \citep{2024Natur.633..318C}, the spatial extent does not deny that they are at $z\sim12$.
If they are $z\sim3$ QGs, the physical sizes become $\sim0.8\,{\rm kpc}$, which is also comparable with the typical size of the QG population at $z=1$--3 \citep[e.g.,][]{2024ApJ...967L..23C}.

%
%
%
%
%
%
\section{Intepretation of the red UV slopes at $\lowercase{z}\sim12$}\label{sec:discussion}
In this section, we discuss the possible origins of red colors in galaxy candidates at $z\sim12$. 
There are four factors that can redden galaxy SEDs in the rest-frame UV: old stellar ages, high stellar metallicities, dust attenuation, and nebular continuum. 
As the age of the Universe at $z\sim12$ is just 350\,Myr, age or metallicity effects can be negligible, reaching at most $\beta_{\rm UV}\sim-2.0$ \citep[e.g.,][]{1999ApJS..123....3L,2014A&A...566A..19C}.
Hereafter, we consider the remaining two possibilities: dust attenuation and strong nebular continuum.
The basic SED fits using \texttt{Prospector}, \texttt{Bagpipes}, and \texttt{eazy} do not produce a nebular-dominated SED: \texttt{Bagpipes} and \texttt{eazy} explain the red color by dust extinction, and \texttt{Prospector} infers a high stellar metallicity of $Z\sim0.5$--$1\,Z_{\odot}$ which is not realistic at $z>10$ \citep{2024ApJ...973....8H,2024ApJ...972..143C,2025NatAs...9..155Z,2025A&A...695A.250A}.

\subsection{case 1: early dust growth?}\label{subsec:dusty}
The simple explanation for their red color is dust extinction.
The significant dust emission/extinction at $z\lesssim8$ is confirmed from many previous ALMA/JWST observations \citep[e.g.,][]{2017MNRAS.466..138K,2019ApJ...874...27T,2020A&A...643A...2B,2020A&A...643A...4F,2022MNRAS.515.3126I,2022MNRAS.512...58F,2022MNRAS.512..989D,2022MNRAS.515.1751W,2023MNRAS.523.3119W,2023Natur.621..267W,2024ApJ...971..161M,2024arXiv241023959B}.
From the \texttt{Bagpipes} SED fitting assuming delayed-$\tau$+burst SFH (Section \ref{sec:data}), the extinction values of the two galaxies are $A_V=0.86_{-0.11}^{+0.12}$ and $0.85_{-0.17}^{+0.21}$ for ID35653 and ID35654, respectively.
The mass-weighted age of two galaxies are $t_{\rm age}=250^{+70}_{-80}\,{\rm Myr}$ and $250^{+60}_{-90}\,{\rm Myr}$, for ID35653 and ID35654, respectively.
This may suggest relatively old, mature stellar populations, though further observations of age-sensitive features such as emission lines or Balmer breaks are required to derive reliable ages.
To confirm that the derived $A_V$ values do not strongly depend on the SFH, we tested a very young case ($t_{\rm age}=5\,{\rm Myr}$) by assuming a constant SFH lasting 10\,Myr.
The change of the $A_V$ values is less than 10\% for both galaxies.

Such large $A_V$ values at $z>10$ are not expected in the latest simulations \citep{2025ApJ...982....7N,2025A&A...696A.157M} and observations \citep{2023MNRAS.520...14C,2024MNRAS.531..997C,2024arXiv240410751A,2025arXiv250119384M,2025A&A...694A.286F}.
\citet{2025ApJ...982....7N} imply the median $A_V\sim0.04\,{\rm mag}$ at $z\sim12$ in their {\sc smuggle} galaxy model, and \citet{2025A&A...694A.286F} reported $A_V\sim0.12\,{\rm mag}$ among 15 galaxies at $z_{\rm spec}>10$.
To compare these galaxies with the previous observations, we convert $A_V$ into the dust mass $M_{\rm dust}$ following \citet{2025A&A...694A.286F}: $M_{\rm dust}\sim2.2\times10^4 A_V(r_d/100{\rm pc})^2$.
Here we utilize the effective radii in F444W derived from {\sc pysersic} \citep[][Zhang et al. in prep]{2024arXiv241206957M} as a proxy of the dust effective radii ($r_d$).

The resulting $M_{\rm dust}$ are $\log M_{\rm dust}[M_{\odot}]=5.28_{-0.15}^{+0.12}$ and $5.21_{-0.16}^{+0.06}$ for ID35653 and 35654, respectively.
In Figure \ref{fig:Mdust_Mstar}, we compare the derived stellar mass and dust mass with those from previous observations summarized in \citet{2025A&A...694A.286F} and simulation \citep{2025ApJ...982....7N}.
Note that shallower extinction curves than Calzetti are expected at high-$z$ \citep[][see also, \citealt{2025arXiv250901795S}]{2025arXiv250214031M,2025arXiv250918266N}, requiring even larger $A_V$ or $M_{\rm dust}$ to reproduce their $\beta_{\rm UV}$.
For the conservative estimation of the stellar mass, the mean value obtained from two different \texttt{Bagpipes} fittings (the delayed-$\tau$ SFH and 10\,Myr constant SFH) and their differences are used as the best estimation and the uncertainty, respectively.
We confirmed that the mean value is consistent with the stellar mass estimated by \texttt{Prospector}.
The estimated $M_{\rm dust}$/$M_{\ast}$ are consistent with the model predictions at $z\sim12$ \citep{2025ApJ...982....7N}.
Also, \citet{2025A&A...696A.157M} shows consistent $M_{\ast}$--$M_{\rm dust}$, but expected blue UV slopes of $\beta_{\rm UV}\sim-2.5$ at $z\sim12$ (i.e., less extinction at the given dust mass).

As suggested in \citet{2025ApJ...982....7N}, grain size distribution is one important factor to explain large $A_V$ at $z\sim12$.
Even though the total dust mass does not change significantly, grain-grain shattering shifts the grain size distribution toward smaller dust grains, and the effective extinction at rest-frame UV-to-optical increases.
If the two galaxies truly exhibit a large extinction of $A_V\sim0.8\,{\rm mag}$, it may suggest early dust shattering due to largely turbulent gas.
Another important factor is the dust distribution.
The effective extinction increases if the dust is very concentrated around star formation sites.
For example, in the blue monstars model \citep{2023MNRAS.522.3986F}, the intensive star formation occurs in compact, dust-embedded SF regions before being affected by radiation-driven outflow, in line with the compact sizes ($\sim300\,{\rm pc}$).

While it is difficult to explicitly say which is the important factor to explain the large extinction at the given dust mass, the two galaxies may be outliers as compared to the ``average'' non-attenuated population at $z>10$.
This is supported in the comparison between other $z>10$ galaxies: they exhibit $\log M_{\rm dust}/M_{\ast}\lesssim-4$, \citet{2025A&A...694A.286F}, while two red galaxies show $\log M_{\rm dust}/M_{\ast}\sim-3$.
In the statistical comparison of $\beta_{\rm UV}$ across the redshift \citep{2023MNRAS.520...14C,2024MNRAS.531..997C,2024arXiv240410751A}, they found $\beta_{\rm UV}$ is getting bluer at higher-$z$, and no dusty galaxies have been identified spectroscopically yet.
The spectroscopic identification of such red, dusty galaxies enables us to demonstrate such a potential mechanism to make ``blue monsters'', and could be a key to understanding abundant bright galaxies at $z>10$.
Furthermore, investigation of the dust amount and extinction at $z>10$ galaxies is important to capture the very early stage of the dust production/growth \citep[e.g.,][]{2025arXiv250214031M,2025arXiv250119384M}.

%
%
%
%
%
%
\label{fig:Mdust_Mstar}
%
%
%
%
%
%

%
%
%
%
%
%
\begin{figure}[t]
\begin{center}
\includegraphics[width=8.8cm,bb=0 0 1000 650, trim=0 1 0 0cm]{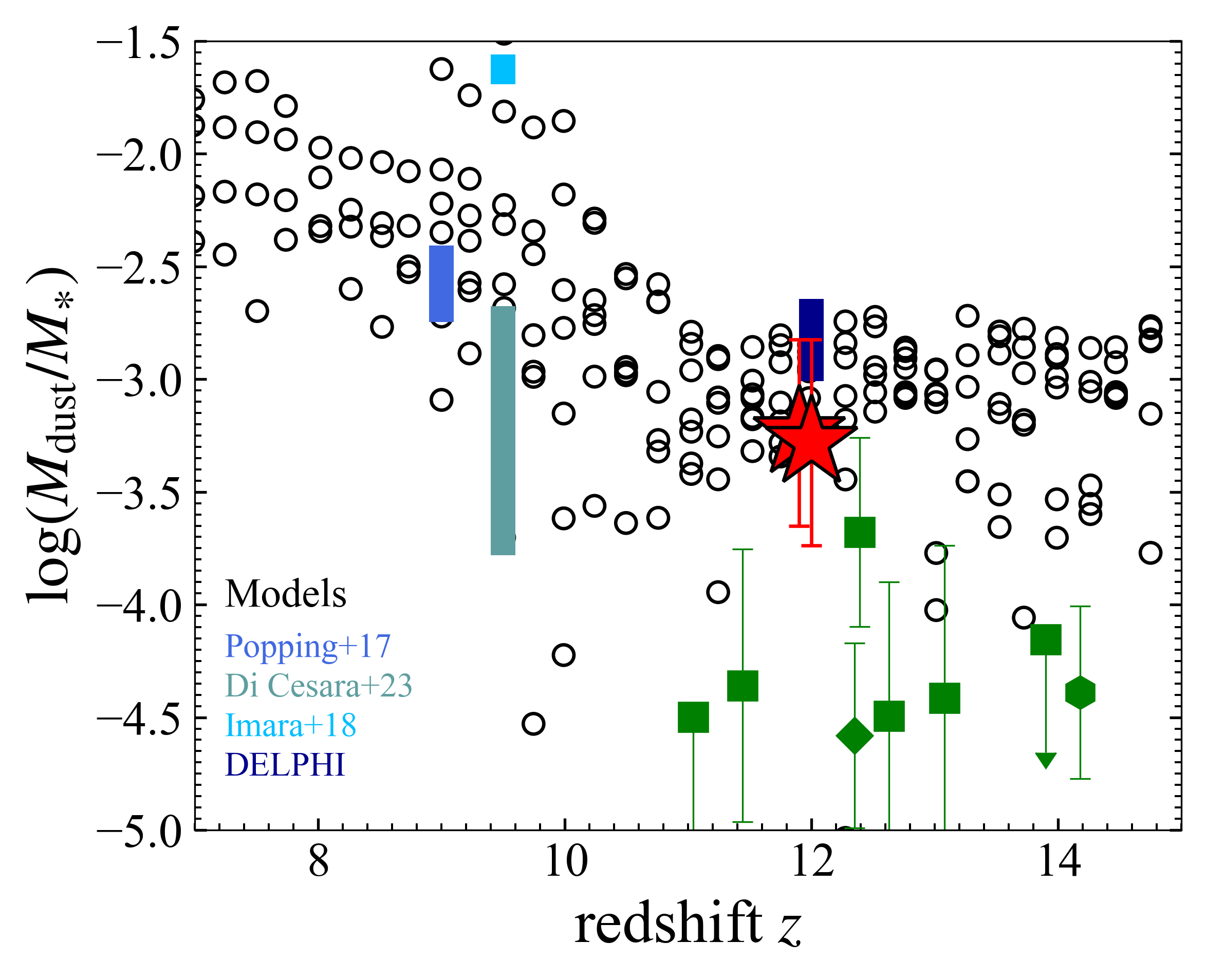}
\caption{The redshift vs the ratio between dust mass ($M_{\rm dust}$) and the stellar mass ($M_{\rm ast}$) for our two $z\sim12$ candidates (red stars), where $M_{\rm dust}$ is derived from $r_{\rm dust}$ and $A_V$. 
The expectation from several models \citep{2017MNRAS.471.3152P,2018ApJ...854...36I,2023MNRAS.519.4632D,2023MNRAS.526.2196M,2025A&A...696A.157M,2025arXiv250918266N} and the constraints from observations \citet{2025A&A...694A.286F} (at $\log M_{\ast}[M_{\odot}]\sim8$--9 when stellar mass estimation is available) are shown.
If the red $\beta_{\rm UV}$ originates from the dust reddening,  the dust-to-stellar mass ratio of the two $z\sim12$ candidates is consistent with the simulations, and larger than the other sources at $z>10$.
} 
\label{fig:Mdust_Mstar}
\end{center}
\end{figure}

\subsection{case 2: nebular continuum dominated galaxy?}\label{subsec:nebular}

%
%
%
%
%
%
\begin{figure*}[htbp]
\begin{center}
\includegraphics[width=18cm,bb=0 0 1000 650, trim=0 1 0 0cm]{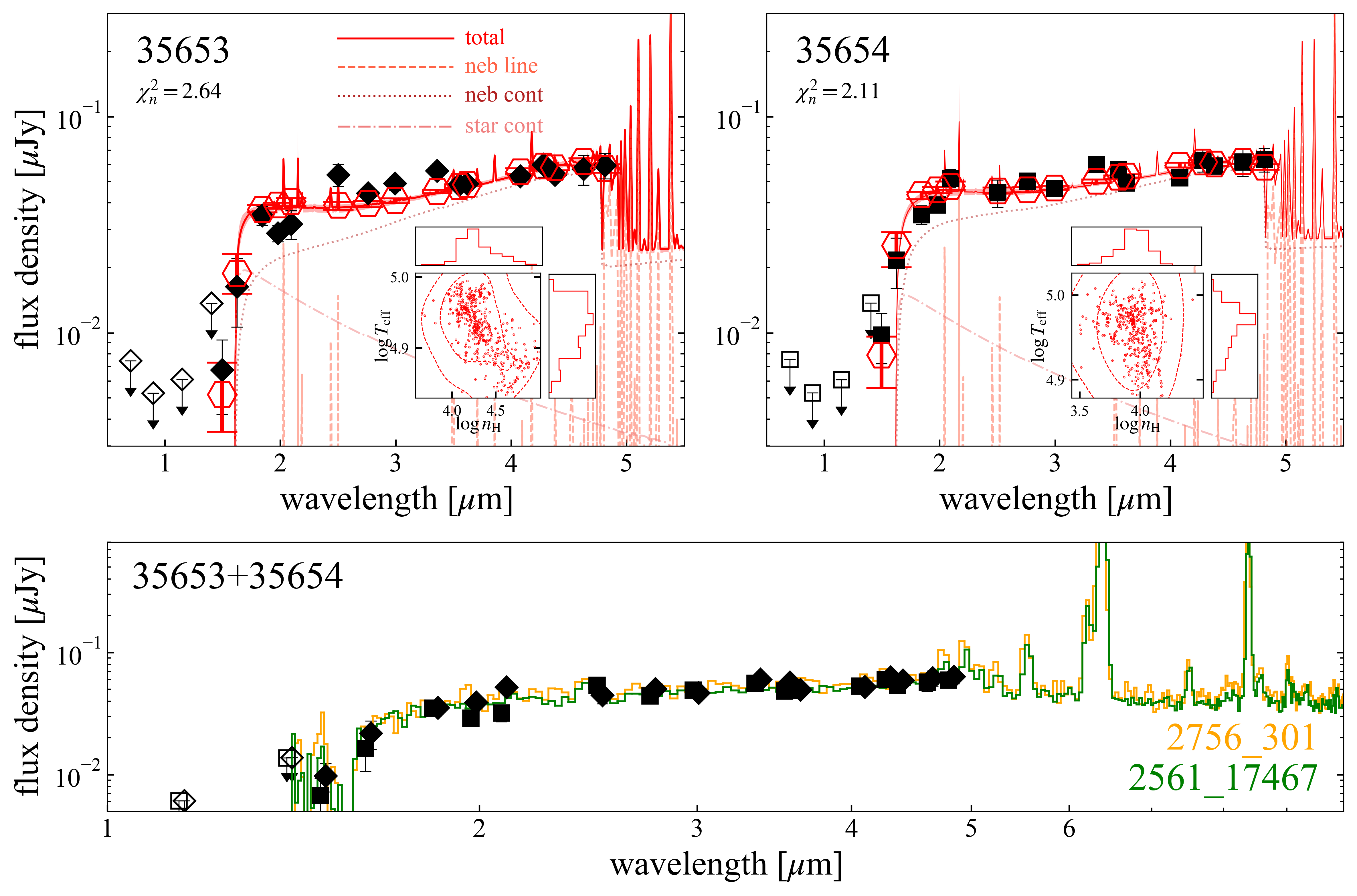}
\caption{(top) The best-fit SED based on \texttt{Cloudy}, with the separated contribution from each component (nebular line, nebular continuum, and stellar/incident radiation). The IDs and $\chi^2_{\rm n}$ values are shown in the top left. The inserted panels show the posterior distribution of the $T_{\rm eff}$ and $n_{\rm H}$, where $T_{\rm eff}$ and $n_{\rm H}$ are the stellar effective temperature of the incident radiation source and gas density surrounding the radiation source. The large $T_{\rm eff}$ and $n_{\rm H}$ are necessary to reproduce red $\beta_{\rm UV}$ from a nebular continuum dominated case. (bottom) Two example PRISM spectra of the nebular continuum-dominated galaxy at $z\sim4$ \citep{2024arXiv240803189K} scaled to two red galaxies. The photometry aligns well with the spectra.}
\label{fig:SED_nebular}
\end{center}
\end{figure*}

Another possible explanation for the red $\beta_{\rm UV}$ at the early epoch is significant nebular continuum contributions to the rest-frame UV.
As predicted in \citet{2024arXiv240803189K}, nebular continuum can potentially make $\beta_{\rm UV}$ values as red as $\beta_{\rm UV}\sim-1.3$ depending on the stellar age ($t_{\rm age}$), gas density ($n_{\rm H}$) and stellar effective temperature ($T_{\rm eff}$).
\citet{2024arXiv241114532S} statistically investigated the UV slope $\beta_{\rm UV}$ of $z>9.5$ galaxies with NIRSpec/PRISM data and suggests that nebular continuum reddening may play a significant role at $z>9.5$.

We explore the physical conditions that would be necessary to explain our observed photometry through strong nebular continuum reddening.
Here we utilized \texttt{Cloudy} version C23 \citep{2023RMxAA..59..327C} with the metal-poor, massive stellar model from \citet{2023AJ....165....2L}.
We adopted IGM absorption from \citet{2014MNRAS.442.1805I} and ISM/CGM absorption from \citet{2025ApJ...983L...2A}.
We fixed the ionization parameter to $\log U_{\rm ion}=-2.0$, metallicity to $\log Z_{\rm gas}[Z_{\odot}]=-2.0$ or $-3.0$ and $t_{\rm age}=0$ (instantaneous burst), and set $n_{\rm H}$ and $T_{\rm eff}$ to free parameters in \texttt{Cloudy} run.
Then we obtained model SEDs, including the radiation from the nebular and incident radiation.
After creating model SEDs, we fit them to the available photometry with additional free parameters of redshift $z$ and absolute scaling.
We utilized {\sc emcee} with 750 iterations in the fitting procedure, and confirmed that the fitting converged.

Figure \ref{fig:SED_nebular} shows the best-fit SEDs with $\log Z_{\rm gas}\,[Z_{\odot}]=-3.0$.
There is no significant change if we assume $\log Z_{\rm gas}\,[Z_{\odot}]=-2.0$.
The best-fit model SEDs infer gas density of $\log n_{\rm H}\,[{\rm cm}^{-2}]\sim4.0$ and effective stellar temperature of $\log T_{\rm eff}\,[{\rm K}]\sim4.95$.
We also confirm that the other stellar incident radiation (e.g., BPASS) can not reproduce $\beta_{\rm UV}>-1.7$ even under high $n_{\rm H}$.
Also, the ionization parameter $U_{\rm ion}$ only has a minor impact on resulting SEDs.

Such $n_{\rm H}$ and $T_{\rm eff}$ are much higher than the typical nebular conditions assumed at high-$z$ \citep[e.g., $n_e\sim10^{-3}$ at $z\sim9.5$,][]{2024MNRAS.533.2488M,2024ApJ...973...47A}. 
The red UV slopes are formed by the dominant free-bound emission and suppressed two-photon emission due to very high gas density \citep{2006agna.book.....O}.
The electron density $n_e$, usually comparable with gas density $n_{\rm H}$ around the H\,{\sc ii} regions, is $10^{2}$--$10^{3}$ \citep[e.g.,][]{2021ApJ...909...78D,2023ApJ...956..139I} while it may depend on ISM phases wihtin galaxy \citep[see also,][]{2025MNRAS.541.1707T,2025arXiv250509186H}.
The stellar temperature reaches $\sim40000\,{\rm K}$ in O-type stars \citep[e.g.,][]{2005A&A...436.1049M}, but it is difficult to explain $T_{\rm eff}\sim100000\,{\rm K}$ in usual initial mass functions (IMF).
The ionizing photon production rate ($\xi_{\rm ion}$) under nebular continuum case is $\log \xi_{\rm ion}\,[{\rm Hz},{\rm erg}^{-1}]\sim26.1$--$26.2$, exceeding those in the typical IMF \citep[e.g.,][]{2018ApJ...855...42S}.
These $\xi_{\rm ion}$ are comparable with those in \citet{2025arXiv250511263N}, suggesting that emission lines from highly excited metal will be detected if $Z>0.05\,Z_{\odot}$.
Also, $\xi_{\rm ion}$ is similar to those of the spectroscopically confirmed nebular continuum-dominated galaxy at $z=4$ \citep{2024arXiv240803189K}, where a strong manifestation may separate a continuum source from the nebula \citep[``offset nebular'' scinario in][]{2024MNRAS.534..523C}.
We show the PRISM spectra of two nebular continuum-dominated $z=4$ galaxies scaled to two red galaxies in Figure \ref{fig:SED_nebular}, which align well with our observed photometry.

We also computed the total hydrogen ionizing photon luminosity ($Q$) and estimated the required mass of the cluster Wolf–Rayet or metal-poor stars following \citet{2024MNRAS.534..523C}.
We found $Q\sim8.0$--$8.7\times10^{54}\,{\rm s}^{-1}$, and $\sim40000$ $Z=0.07\,Z_{\odot}$ Wolf–Rayet or $\sim90000$ metal-poor stars with $\sim100\,M_{\odot}$ are needed to account for this ionizing photon luminosity.
The expected stellar mass reaches $\sim10^7\,M_{\odot}$, which is too massive for Pop-{\sc iii} star clusters \citep[e.g.,][]{2021MNRAS.508.3226A}.
The absolute UV magnitude of these galaxies ($M_{\rm UV}\sim-19\,{\rm mag}$) also may be too bright for the Pop-{\sc iii} candidates, as it is challenging to form such massive Pop-{\sc iii} only star clusters \citep[e.g.,][see also, \citealt{2025arXiv250111678F}]{2019MNRAS.488.2202J,2024MNRAS.527.5102V} and therefore much higher metallicity than those assumed here is expected at this $M_{\rm UV}$ regime \citep[$\log Z_{\rm gas}>-2.0$,][]{2025A&A...696A.157M}.

Another possibility to explain the red UV slopes is the presence of the AGN.
AGNs can power surrounding nebular gas instead of the stars, and produce red UV slopes \citep{2022ApJ...931L..25I,2022MNRAS.513.5134N}.
The expected colors from these AGN SEDs are consistent with those of two red galaxies.
However, since these galaxies exhibit an extended morphology, this is unexpected from the nebula nearby the BH at $\lesssim100\,{\rm pc}$.

\subsection{case 1 + case 2, combined}\label{subsec:combine}
Is it possible that the red $\beta_{\rm UV}$ represents a combination of the two cases described above?
Case 1 (Section \ref{subsec:dusty}) implies relatively metal-enriched and mature galaxies, while case 2 implies metal-poor and extremely young galaxies (Section \ref{subsec:nebular}).
Therefore, these two cases represent completely opposite natures, and the composite scenario is unlikely.
Such a composite scenario would likely require multi-episode star formation with both a mature, dust-reddened component and a young, metal-poor component.
If both components are comparably red and bright, it is difficult to separate the contributions from these components from observations \citep[c.f.,][]{2023ApJ...948..126G}.
It is not required that both components be red; a configuration with red and blue components, for example, dusty red ($\beta_{\rm UV}>-1.3$) and young blue ($\beta_{\rm UV}<-1.5$), is also acceptable.
Since many more red slopes are dispreferred from either dusty or nebular cases (Figure \ref{fig:Mdust_Mstar} and \ref{fig:SED_nebular}, see also \citealt{2024arXiv240803189K}), one of these dominates within a galaxy when the red and blue components are mixed. 

%
%
%
%
%
%
\section{Prospects and summary}\label{sec:summary}
In this paper, we report the discovery of two red galaxy candidates at $z\sim12$ based on the full 20 NIRCam band coverage from the combination of the UNCOVER \citep{2024ApJ...974...92B} and MegaScience \citep{2024ApJ...976..101S} surveys.
We found seven robust $z>10$ candidates from \texttt{Prospector} fitting, where two of them had large changes of the $z_{\rm phot}$ owing to the effective addition of the medium bands.
The $M_{\rm UV}$ and $\beta_{\rm UV}$ relationship aligns well with those from \citep{2025arXiv250703124A}, where they utilized a similar photometric dataset including multiple medium bands and suggest systematically redder $\beta_{\rm UV}$ than the $M_{\rm UV}$--$\beta_{\rm UV}$ relation from broad band-based estimations.
This alignment supports a potential bias toward the intrinsically blue galaxies in a broad-band-based LBG selection.
We identified two red galaxies which are significant outliers from any published $M_{\rm UV}$--$\beta_{\rm UV}$ relations with $M_{\rm UV}\sim-19\,{\rm mag}$ and $\beta_{\rm UV}\gtrsim-1.5$.
Their $p(z)$ suggest $z\sim12$ solution is highly plausible with $p(z<7)\lesssim5\,\%$.

The red rest-frame UV colors ($\beta_{\rm UV}\gtrsim-1.5$) can be explained by the reddening due to strong dust extinction ($A_V\sim0.8$) or nebular continuum contribution from dense gas surrounding the hot stars ($\log n_{\rm H}\,[{\rm cm}^{-2}]\sim4.0$ and $\log T_{\rm eff}\,[{\rm K}]\sim4.95$).
While it is very difficult to completely rule out a low-$z$ solution, the effective addition of medium bands is very critical to select high-$z$ galaxies.
Since the selection based on fewer photometric points (i.e., broadband) leads to contamination of low-$z$ strong line emitters \citep[e.g.,][]{2023Natur.622..707A} or to biases towards blue galaxies \citep[e.g.,][]{2014ApJ...793..115B}, photometric surveys covering the key SED representation bring us an unbiased view on the early Universe.

Red galaxies at $z>10$ are interesting from the perspective of not only the metal-enriched galaxies with early dust growth, but also the extremely young, potentially metal-poor galaxies.
Future spectroscopic confirmation and further detailed modeling in the rest-frame UV/optical wavelength, especially with Balmer lines, will confirm the nature of the red galaxies in the early Universe and provide new insight into early galaxy formation, which may have been missed so far.

\acknowledgments

I.M. and K.A.S. thank Shelly Meyett, the MegaScience Program Coordinator, for invaluable assistance in designing WOPR 88967 and ensuring that this program was fully observed. 
This proposal was conceived of and developed at the International Space Science Institute (ISSI) in Bern, through ISSI International Team project \#562. 
This work is based on observations made with the NASA/ESA/CSA James Webb Space Telescope. 
The raw data were obtained from the Mikulski Archive for Space Telescopes at the Space Telescope Science Institute, which is operated by the Association of Universities for Research in Astronomy, Inc., under NASA contract NAS 5-03127 for JWST. 
These observations are associated with JWST Cycle 2 GO program \#4111, and this project has gratefully made use of a large number of public JWST programs in the A2744 field, including JWST-GO-2641, JWST-ERS-1324, JWST-DD-2756, JWST-GO-2883, JWSTGO-3538, and JWST-GO-3516. Support for program JWSTGO-4111 was provided by NASA through a grant from the Space Telescope Science Institute, which is operated by the Association of Universities for Research in Astronomy, Incorporated, under NASA contract NAS5-26555.
I.M. acknowledges funding from JWST-GO-04111.035.
P.D. warmly acknowledges support from an NSERC discovery grant (RGPIN-2025-06182).



%
%
%
%
%
%
%
%
%
%
%
%

\clearpage
\bibliography{main.bib}{}
\bibliographystyle{apj.bst}

\end{document}